\documentclass[twocolumn,letter]{jpsj3}
\usepackage{txfonts}
\title{Unconventional Superconductivity in Heavy Fermion UTe$_2$}

\author{
Dai~Aoki$^{1,2}$\thanks{E-mail: aoki@imr.tohoku.ac.jp}, 
Ai~Nakamura$^1$,
Fuminori~Honda$^1$,
DeXin~Li$^1$,
Yoshiya~Homma$^1$
Yusei~Shimizu$^1$,
Yoshiki~J.~Sato$^1$,
Georg~Knebel$^2$,
Jean-Pascal~Brison$^2$,
Alexandre~Pourret$^2$,
Daniel~Braithwaite$^2$, 
Gerard~Lapertot$^2$,
Qun~Niu$^2$,
Michal Vali\v{s}ka$^2$,
Hisatomo~Harima$^3$, and
Jacques~Flouquet$^2$
}

\inst{%
$^1$IMR, Tohoku University, Oarai, Ibaraki, 311-1313, Japan\\
$^2$University Grenoble Alpes, CEA, IRIG-PHELIQS, F-38000 Grenoble, France\\
$^3$Graduate School of Science, Kobe University, Kobe 657-8501, Japan
}

\abst{%
We grew single crystals of the recently discovered heavy fermion superconductor UTe$_2$, and measured the resistivity, specific heat and magnetoresistance.
Superconductivity (SC) was clearly detected at $T_{\rm sc}=1.65\,{\rm K}$ as sharp drop of the resistivity in a high quality sample of $\mbox{RRR}=35$.
The specific heat shows a large jump at $T_{\rm sc}$ indicating strong coupling.
The large Sommerfeld coefficient, $\gamma=117\,{\rm mJ K^{-2}mol^{-1}}$ extrapolated in the normal state
and the temperature dependence of $C/T$ below $T_{\rm sc}$ are the signature of unconventional SC.
The discrepancy in the entropy balance at $T_{\rm sc}$ between SC and normal states points out that
hidden features must occur.
Surprisingly, a large residual value of the Sommerfeld coefficient seems quite robust ($\gamma_0 /\gamma \sim 0.5$).
The large upper critical field $H_{\rm c2}$ along the three principal axes favors spin-triplet SC.
For $H\parallel b$-axis, our experiments do not reproduce the huge upturn of $H_{\rm c2}$ reported previously.
This discrepancy may reflect that $H_{\rm c2}$ is very sensitive to the sample quality. 
A new perspective in UTe$_2$ is the proximity of a Kondo semiconducting phase predicted by the LDA band structure calculations.
}

\begin{document}
\maketitle
Unconventional superconductivity (SC) attracts much attention in the strongly correlated electron systems,
in particular, the microscopic coexistence of ferromagnetism and SC~\cite{Aok12_JPSJ_review,Aok19} discovered in UGe$_2$~\cite{Sax00}, URhGe~\cite{Aok01}, and UCoGe~\cite{Huy07}.
One of the highlights is the field-reentrant (-reinforced) SC~\cite{Lev05,Aok09_UCoGe}
in transverse magnetic field with respect to the ferromagnetic (FM) alignment;
when the magnetic field is applied along the hard magnetization axis ($b$-axis) in URhGe and UCoGe,
the Curie temperature $T_{\rm Curie}$ is suppressed. 
The collapse of $T_{\rm Curie}$ enhances the FM fluctuations which boost SC.
In addition Fermi surface (FS) instabilities give an extra source for the enhancement of the upper critical field, $H_{\rm c2}$.~\cite{Yel11,Gou16,Bas16}
Spin-triplet pairing is clearly realized for the three uranium ferromagnets.

Very recently, SC was discovered in the heavy fermion paramagnet UTe$_2$~\cite{Ran19}, 
with a rather high superconducting transition temperature, $T_{\rm sc}=1.6\,{\rm K}$.
Furthermore, from the large $H_{\rm c2}$, exceeding the Pauli limit, and the constant Knight shift through $T_{\rm sc}$, the spin-triplet SC seems to occur.
The great interest compared to the previous cases is that the ground state is paramagnetic at the verge of FM order above the appearance of SC.
In order to study these results in more detail, we grew single crystals and measured the resistivity, specific heat and magnetoresistance at low temperatures.

UTe$_2$ crystallizes in the body-centered orthorhombic structure (UTe$_2$-type) with the space group $Immm$ ({\#}71, $D_{2h}^{25}$). 
The lattice parameters are 
$a=4.165\,{\rm \AA}$,
$b=6.139\,{\rm \AA}$, and
$c=13.979\,{\rm \AA}$. 
The $c$-axis is quite long, but the corresponding Brillouin zone is not very flat because of the body-centered orthorhombic structure.
A paramagnetic ground state and heavy electronic states with a large Sommerfeld coefficient $\gamma\sim 120$--$150\,{\rm mJ K^{-2}mol^{-1}}$ have been reported.~\cite{Ran19,Ike06_UTe2}
The magnetic susceptibility shows Curie-Weiss behavior above $150\,{\rm K}$ with effective moments close to the 5$f^2$ or 5$f^3$ free ion value, indicating a 5$f$-localized nature at high temperatures.
Note that the Weiss temperatures obtained from fits above $150\,{\rm K}$ for $a$, $b$, and $c$-axes are negative, suggesting antiferromagnetic interactions at high temperatures. 
For $H\parallel b$-axis, the susceptibility shows a broad maximum with $T_{\chi_{\rm max}}\sim 35\,{\rm K}$.
The magnetization curve at $2\,{\rm K}$ is anisotropic.
The easy-axis ($a$-axis) magnetization reaches $0.5\,\mu_{\rm B}/{\rm U}$ at $7\,{\rm T}$,
while the hard-axes ($c$ and $b$-axes) magnetizations are only $0.15$ and $0.09\,\mu_{\rm B}/{\rm U}$, respectively at $7\,{\rm T}$.~\cite{Ran19,Ike06_UTe2}

Single crystals of UTe$_2$ were grown using chemical vapor transport (CVT) method with Iodine as transport agent, in Oarai and in Grenoble, as described in Ref.~\citen{Ran19}.
The self-flux method was also used in Oarai. 
The starting materials with the ratio, U : Te = 22 : 78 (at{\%}) were put into an alumina crucible in a double sealed (tantalum/quartz) ampoule.
The Te-flux was removed by spinning off in a centrifuge. 
The single crystals obtained from both CVT and flux methods were checked by single crystal X-ray analysis. 
The lattice parameters and the atomic coordinates are well-defined, in good agreement with the previous results.~\cite{Ike06_UTe2}
The resistivity at zero field down to $0.1\,{\rm K}$ was measured by the four-probe AC method.
The specific heat was measured by the relaxation method at temperatures down to $0.4\,{\rm K}$ in PPMS,
and at lower temperatures down to $0.1\,{\rm K}$ in a dilution refrigerator using a homemade calorimetric cell.
The magnetoresistance was measured by the four-probe DC method at low temperatures down to $0.03\,{\rm K}$, 
and at high fields up to $15\,{\rm T}$
in a top-loading dilution refrigerator. 
The sample with the current along the $a$-axis was rotated for the field directions from $c$ to $b$ and from $b$ to $a$.
A homemade dilution refrigerator with a $16\,{\rm T}$ magnet was also used in Grenoble.

\begin{figure}[tbh]
\begin{center}
\includegraphics[width=0.8\hsize,pagebox=artbox]{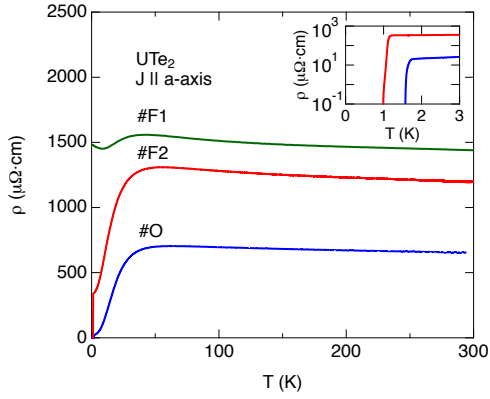}
\end{center}
\caption{(Color online) Temperature dependence of the resistivity for the current along $a$-axis in different samples of UTe$_2$. 
The samples grown using the flux method are denoted by {\#}F1 and {\#}F2. The sample grown using the CVT method is denoted by {\#}O. The inset shows the resistivity at low temperatures for {\#}F2 and {\#}O with the logarithmic scale of vertical axis.}
\label{fig:resist}
\end{figure}
Figure~\ref{fig:resist} shows the temperature dependence of the resistivity for the current along the $a$-axis of three different samples.
The resistivity for the CVT sample ({\#}O) slightly increases linearly down to $50\,{\rm K}$ and decreases rapidly with further decreasing temperature,
indicating typical heavy fermion behavior.
At low temperature below $\sim 4 \,{\rm K}$, the resistivity follows a $T^2$ dependence, indicating Fermi liquid nature,
and becomes zero, revealing the superconducting transition at $T_{\rm sc}=1.65\,{\rm K}$ defined by the mid point of the resistivity drop.
The residual resistivity $\rho_0$ and the residual resistivity ratio RRR are $18.5\,\mu\Omega\!\cdot\!{\rm cm}$ and $35$, respectively.
The $A$ coefficient for the $T^2$ dependence is $0.88\,\mu\Omega\!\cdot\!{\rm cm}/{\rm K^2}$.

Interestingly, the lower quality sample denoted by {\#}F2 ($\mbox{RRR}=4$ and $\rho_0=340\,\mu\Omega\!\cdot\!{\rm cm}$) also shows the superconducting transition at $T_{\rm sc}=1.1\,{\rm K}$. 
The resistivity at room temperature is nearly twice larger than that for {\#}O.
The lowest quality sample ({\#}F1) does not show the superconducting transition down to $0.1\,{\rm K}$,
instead a small upturn appears. 

\begin{figure}[tbh]
\begin{center}
\includegraphics[width=1 \hsize,pagebox=cropbox,clip]{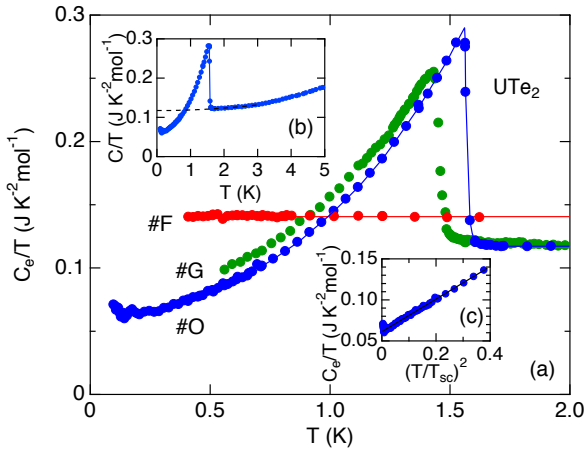}
\end{center}
\caption{(Color online) (a) Electronic specific heat in the form of $C_{\rm e}/T$ vs $T$ for the CVT ({\#}O, {\#}G) and flux ({\#}F) grown samples of UTe$_2$. $T_{\rm sc}$ is $1.57\,(1.46)\,{\rm K}$ for {\#O} ({\#}G) sample.
The small upturn below $0.12\,{\rm K}$ is most likely due to the nuclear contribution of Te or impurities.
(b) The total specific heat in sample {\#}O. The dotted line is the fitting in the normal state to subtract the phonon contribution.
(c) The low temperature part of $C_{\rm e}/T$ in sample {\#}O in the form of $C_{\rm e}/T$ vs $(T/T_{\rm sc})^2$.
}
\label{fig:Cp}
\end{figure}
Figure~\ref{fig:Cp}(a) shows the electronic specific heat down to $0.1\,{\rm K}$ for two CVT ({\#}O, {\#}G) and flux ({\#}F) grown samples.
A $T^3$ phonon contribution is subtracted by fitting the normal state, as shown in Fig.~\ref{fig:Cp}(b).
The Sommerfeld coefficient ($\gamma$-value) is $\gamma = 117\,{\rm mJ\,K^{-2}mol^{-1}}$, indicating the presence of heavy electronic states in UTe$_2$.
Bulk SC is clearly demonstrated by the large and sharp specific heat jump 
with $\Delta C_{\rm e}/\gamma T_{\rm sc}=1.51$ and $T_{\rm sc}=1.57\,{\rm K}$ for {\#}O sample,
compared to the weak coupling BCS value ($\Delta C_{\rm e}/\gamma T_{\rm sc}=1.43$).
At low temperatures, $C_{\rm e}$ varies as $\gamma_0 T + \eta T^3$ as shown in Fig.~\ref{fig:Cp}(c) for sample {\#}O.
These results together with the field dependence of $C/T$~\cite{suppl} are not at all consistent with that for the conventional BCS type.

Remarkably, a large residual $\gamma$-value ($\gamma_0\sim 61\,{\rm mJ\,K^{-2}mol^{-1}}$), 
which is equivalent to $\sim 50\,{\%}$ of the normal state $\gamma$-value, 
is observed consistently on all SC samples for $T \to 0\,{\rm K}$, 
despite the fact that a sharp specific heat jump is detected at $T_{\rm sc}$. 
A large residual $\gamma$-value, $\gamma_0$, is observed in FM superconductors,
with $\gamma_0$ proportional to the square root of the ordered moment, $M_0$, ($\gamma_0\propto M_0^{1/2}$).~\cite{Aok19}
If UTe$_2$ is a paramagnet down to $0\,{\rm K}$, 
such a large residual $\gamma$-value is not expected. 

Compared to the previous report~\cite{Ran19}, both the data in {\#}O and {\#G} samples confirm that the residual term remains always very close to $50\,{\%}$ of the normal state $\gamma$-value. 
The explanation for this term~\cite{Ran19} is that only half the FS would be paired, as it happens in a short temperature range close to $T_{\rm sc}$ in superfluid $^3$He under field, in the so-called A1 phase~\cite{LeggettRMP1975}.
Theoretically, in the case of large spin-orbit coupling, a non-unitary state is expected in FM superconductors (due to band polarization), but it is excluded in orthorhombic paramagnetic superconductors left with only one dimensional representation~\cite{MineevBook} (they are possible only in the case of multiple second order transitions or of first order transition). 
However, the fact that $H_{\rm c2}$ in UTe$_2$ overcomes the Pauli limitation for all field directions supports the hypothesis that the $d$-vector can re-orient under magnetic field, so that it should be considered in the weak spin-orbit coupling regime.
Indeed, the alternative explanation for the absence of Pauli limitation along the hard axes, that the external field remains smaller than the internal exchange field \cite{MineevPRB2010} cannot apply in paramagnetic systems.

For such a weak spin-orbit coupling regime, multidimensional representations are allowed and non-unitary states are possible in paramagnetic materials~\cite{HillierPRL2012}. 
These non-unitary states are extreme ``equal spin pairing'' (ESP) states, with only one of the two spin directions paired ($d$-vector of the form $\mathbf{d(k)}=\varphi(\mathbf{k})(1,i,0)$). 
So in UTe$_2$, in contrast to the A1 phase of superfluid $^3$He, this non-unitary state, where only half the FS is paired, would extend down to $0\,{\rm K}$. 
The proposed mechanism for the stabilization of such an unfavorable state (half the condensation energy is lost) is through a linear coupling to the magnetization of the system~\cite{HillierPRL2012}, strongly boosted here by the proximity to a FM instability. 
The SC transition would then give rise to a subdominant FM order parameter, reinforced by the transfer of spin-down electrons to spin-up state to gain condensation energy~\cite{TakagiPTP1974}. 

If this extreme non-unitary state is confirmed, it also means that the specific heat jump at $T_{\rm sc}$ is particularly large (more than twice the BCS value), as it should be compared to only half the normal state value. 
This definitely means that UTe$_2$ is in a strong coupling regime. 
This justifies the choice, in Ref.~\citen{Ran19}, of the strong coupling constant in zero field $\lambda \approx 0.75$ for the analysis of $H_{\rm c2}$. 


We also find that the entropy balance reveals that the normal state specific heat does not have a strictly constant $\gamma$-value;
the entropy at $T_{\rm sc}$ is $0.21\,{\rm J\,K^{-1}mol^{-1}}$ from the SC side,
while the entropy at $T_{\rm sc}$ assuming a normal state with constant $C_{\rm e}/T$ is $0.19\,{\rm J\,K^{-1}mol^{-1}}$.
The entropy discrepancy, approximately $10\,{\%}$, may imply a rapid increase of $C_{\rm e}/T$ in the normal state upon cooling due to the development of FM fluctuations.
Note that, $C/T$ at $0.39\,{\rm K}$ reaches $140\,{\rm mJ K^{-2} mol^{-1}}$ at a field close to $H_{\rm c2}$ for $H\parallel c$-axis.~\cite{suppl}
On the other hand, as shown in Fig.~\ref{fig:Cp}(a), the non-superconducting sample ({\#}F) does not show any upturn of $C_{\rm e}/T$ down to $0.4\,{\rm K}$
Its residual $\gamma$-value is about $\sim 140\,{\rm mJ K^{-2}mol^{-1}}$.

\begin{figure}[tbh]
\begin{center}
\includegraphics[width=0.8 \hsize,pagebox=cropbox,clip]{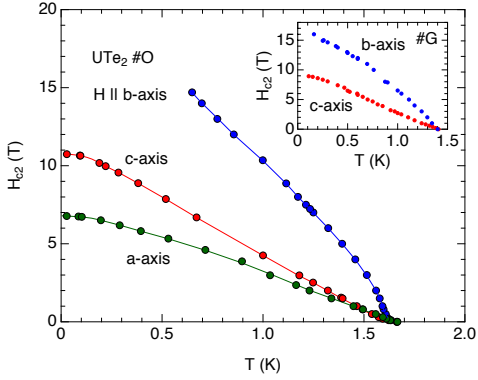}
\end{center}
\caption{(Color online) Temperature dependence of the superconducting upper critical field $H_{\rm c2}$ defined by the mid points of the resistivity drop for $H\parallel a$, $b$, and $c$-axes on sample {\#}O in UTe$_2$.
The inset shows the $H_{\rm c2}$ curves for $H\parallel b$ and $c$-axes defined by zero resistivity on sample {\#}G.}
\label{fig:Hc2}
\end{figure}
The temperature dependence of $H_{\rm c2}$ for $H\parallel a$, $b$, and $c$-axes in {\#}O is shown in Fig.~\ref{fig:Hc2}.
Here the values of $H_{\rm c2}$ and $T_{\rm sc}$ are defined at the midpoint of the resistivity drop from the field/temperature scans.
$H_{\rm c2}$ for $T \to 0\,{\rm K}$ is extremely large. 
For $H\parallel a$ and $c$-axes, the values of $H_{\rm c2}(0)$ are $6.8$ and $10.7\,{\rm T}$, respectively.
For $H\parallel b$-axis, $H_{\rm c2}(0)$ highly exceeds our maximum field $15\,{\rm T}$.
The inset shows the $H_{\rm c2}$ curves defined by zero resistivity on a lower $T_{\rm sc}$ sample {\#}G.

The Pauli limit of $H_{\rm c2}$ at $0\,{\rm K}$ for a singlet superconductor can be estimated as $H_{\rm P}=\sqrt{2}\Delta/(g\mu_{\rm B})=1.84T_{\rm sc}$,
assuming the free-electron value for the $g$-factor, $g=2$, and a weak-coupling regime.
Clearly, $H_{\rm c2}(0)$ in UTe$_2$ violates this paramagnetic limit for all directions.
With strong-coupling effects, the Pauli limit reaches higher values.~\cite{Bul88}
Using, for example, the value of the strong-coupling constant $\lambda=0.75$ reported in Ref.~\citen{Ran19}, it would be possible to reproduce $H_{\rm c2}$ along the $a$-axis with $g < 0.5$.
However, for $H \parallel c$ or $b$, and the same value of $\lambda = 0.75$,
the violation of the Pauli limit is still too strong;
the fitting $H_{\rm c2}$ with the Pauli limit for for these directions would require a vanishing $g$-factor.
Hence, indeed, the $H_{\rm c2}$ curves strongly support the absence of the Pauli limit
and therefore spin triplet SC.

For $H\parallel a$ and $c$-axes, the temperature dependences of $H_{\rm c2}$ are similar to the previous results.~\cite{Ran19}
However, the $H_{\rm c2}$ curve for $H\parallel b$-axis is quite different from that in Ref.~\citen{Ran19} at low temperature, where a strong increase of $H_{\rm c2}$ below $1.2\,{\rm K}$ is shown.
On the other hand, we observe an anomalous linear increase up to $16\,{\rm T}$ at low temperature.

\begin{figure}[tbh]
\begin{center}
\includegraphics[width=1 \hsize,pagebox=cropbox,clip]{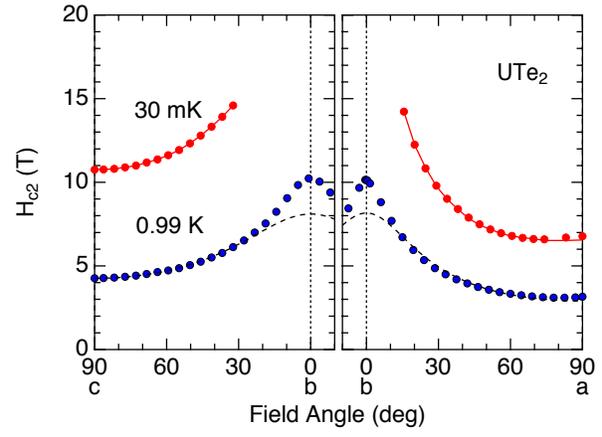}
\end{center}
\caption{(Color online) Angular dependence of $H_{\rm c2}$ at $30\,{\rm mK}$ and $0.99\,{\rm K}$ in UT$_2$ for {\#}O. The dotted lines at $0.99\,{\rm K}$ are the results of fitting by the effective mass model.}
\label{fig:AngHc2}
\end{figure}
Figure~\ref{fig:AngHc2} shows the angular dependence of $H_{\rm c2}$ from $b$ to $c$-axis, and from $b$ to $a$-axis.
$H_{\rm c2}$ for $H\parallel b$-axis at $30\,{\rm mK}$ exceeds the maximum field, $15\,{\rm T}$.
$H_{\rm c2}$ decreases by tilting the field direction from $b$-axis.
The decrease of $H_{\rm c2}$ from $b$ to $a$-axis is more significant than that from $b$ to $c$-axis, as expected from the anisotropy of $H_{\rm c2}$.
At $0.99\,{\rm K}$, $H_{\rm c2}$ for $H\parallel b$-axis is $10.2\,{\rm T}$, revealing the maximum.
The values of $H_{\rm c2}$ for $H\parallel b$-axis determined from $b$ to $a$-axis and from $b$ to $c$-axis are almost identical, indicating that the misorientation for $H\parallel b$-axis is negligibly small.
The acute increase of $H_{\rm c2}$ near $b$-axis cannot be explained by the conventional effective mass model based on the assumption of an ellipsoidal FS, as shown in the dotted lines in Fig.~\ref{fig:AngHc2}.

\begin{figure}[tbh]
\begin{center}
\includegraphics[width=0.8 \hsize,pagebox=cropbox,clip]{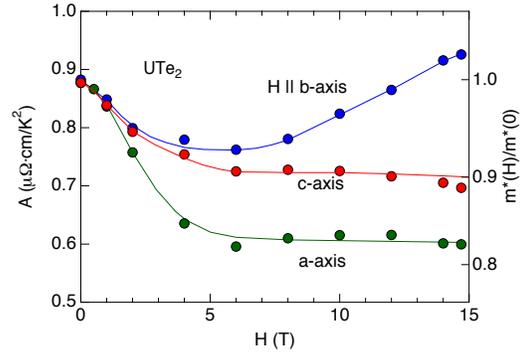}
\end{center}
\caption{(Color online) Field dependence of the resistivity $A$ coefficient for $H \parallel a$, $b$, and $c$-axes in UTe$_2$. The right axis shows the corresponding normalized effective mass, assuming the Kadowaki-Woods relation.}
\label{fig:A_coef}
\end{figure}
In order to capture the field dependence of the effective mass and its consequence on the strong coupling,
the temperature dependence of the resistivity under magnetic field in the temperature range above $T_{\rm sc}$ was measured for $H\parallel a$, $b$, $c$-axes.
The resistivity roughly follows the $T^2$ dependence up to $\sim 3.5\,{\rm K}$ even at high fields.
Thus we extract the resistivity $A$ coefficients, as shown in Fig.~\ref{fig:A_coef}.
The results suggest the field dependence of the effective mass, as indicated in the right axis, based on the assumption of the Kadowaki-Woods relation, $A \propto \gamma^2 \propto {m^\ast}^2$.
The $A$ coefficients decrease initially for $a$, $b$, and $c$-axes, 
however above $5\,{\rm T}$, a clear increase of $A$ is detected only for $H\parallel b$-axis.
The broad maximum in the susceptibility for $H \parallel b$-axis at $T_{\chi_{\rm max}}=35\,{\rm K}$, which remains at least up to $5.5\,{\rm T}$ (not shown), suggests the metamagnetic-like transition at $H_{\rm m}\sim 35\,{\rm T}$~\cite{Aok13_CR,Kna12}
with an enhancement of the magnetic fluctuations associated with the FS change.
It may set the large field-scale for an increase of $m^\ast (H)$ and thus $\lambda (H)$.

\begin{figure}[tbh]
\begin{center}
\includegraphics[width=\hsize,pagebox=cropbox,clip]{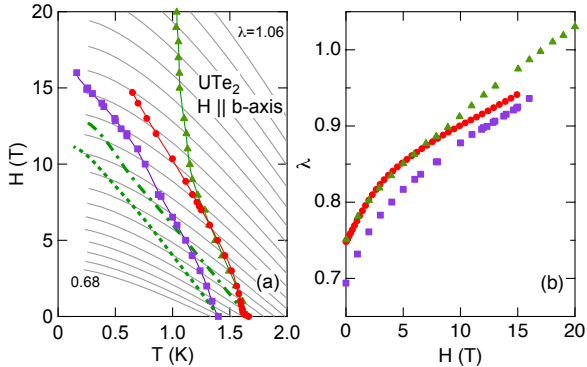}
\end{center}
\caption{(Color online) (a) Comparison of the three data set of $H_{\rm c2}$ for $H\parallel b$ in UTe$_2$ (green triangles~\cite{Ran19}, red circles ({\#O}), and violet squres ({\#}G). 
The gray solid lines are the $H_{\rm c2}$ calculations for $\lambda$ varied between 0.68 and 1.06 by steps of 0.02, used to extract the values of $\lambda (H)$ to reproduce the different curves.
The dashed doted line is the calculated $H_{\rm c2}$ using the same $\lambda (H)$ which reproduces the diverging behavior~\cite{Ran19}, but with an average initial Fermi velocity $15\,{\%}$ larger. 
The dashed line is the calculated $H_{\rm c2}$ using again the same $\lambda (H)$ in Ref.~\citen{Ran19}, 
but with $15\,{\%}$ smaller $T_{\rm sc}(0)$ (characteristic frequency $\Omega$ is $15\,{\%}$ smaller).
(b) Comparison of field dependence of $\lambda$ required to reproduce the $H_{\rm c2}$ curves shown in panel (a), and with the same corresponding symbols. 
Note that although the behavior of $H_{\rm c2}$ shows no divergence in the present work, 
there are only minor differences between the different $\lambda (H)$.
}
\label{fig:lambda}
\end{figure}

Let us comment on the differences observed in the three different results for $H\parallel b$-axis between previous report~\cite{Ran19} and our data ({\#}O and {\#}G). 	
The most evident is the behavior of $H_{\rm c2}$ for $H\parallel b$, which displays a singular divergence only in Ref.~\citen{Ran19}.
At first glance, one could believe that the new data presented here contradict this initial finding. 
However, closer inspection may reconcile all measurements. 
In Fig.~\ref{fig:lambda}(a), we show the experimental $H_{\rm c2}$ data together with the calculations of $H_{\rm c2}$ for different (constant) values of the strong coupling constant $\lambda$. 
These calculations are used to extract the field dependence $\lambda (H)$ required to reproduce the data. 
This is the same procedure as used for UCoGe and URhGe~\cite{WuNatCom2017}, 
and also for UTe$_2$~\cite{Ran19}; 
all calculations are performed for a constant characteristic frequency (field independent) $\Omega=34.3\,{\rm K}$, and a constant band Fermi velocity $v_{\rm F,band}\approx 19200\,{\rm m/s}$. 
In the $H_{\rm c2}$ calculation, this band Fermi velocity is renormalized by $1/(1+\lambda(H))$. 
The value of the band Fermi velocity was fixed to fit the $H_{\rm c2}$ for $H \parallel a$ of Ref.~\citen{Ran19}, (with a field independent $\lambda =0.749$). 
Again, as for UCoGe and as done in Ref.~\citen{Ran19}, we assumed an isotropic Fermi velocity for all field directions, so that the much larger initial slope observed for $H\parallel b$ is explained by the strong initial increase of $\lambda$ with field. 
The field dependence of $\lambda$ for the three data sets are shown in Fig.~\ref{fig:lambda}(b).

First, let us note that although the temperature dependences of $H_{\rm c2}$ in the present work are quite different from the previous results~\cite{Ran19}, the required field dependence of $\lambda$ are very similar for all data sets. 
In particular, $\lambda (H)$ for {\#}G seems just shifted with respect to the previous one~\cite{Ran19}. 
The same physics is at work in all samples. 
By contrast, the diverging behavior of $H_{\rm c2}$ is very sensitive to the exact balance between the orbital limitation of $H_{\rm c2}$ calculated at constant $\lambda$ and the increase of $\lambda (H)$. 

\begin{figure}[bt]
\begin{center}
\includegraphics[width=1 \hsize,pagebox=cropbox,clip]{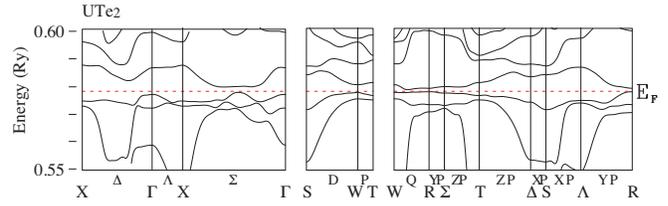}
\end{center}
\caption{(Color online) Band structure calculated by the LDA method near the Fermi energy in UTe$_2$. XP, YP, and ZP lines correspond to $x$, $y$, and $z$-planes with $k_x$, $k_y$, and $k_z=0$, respectively.}
\label{fig:band}
\end{figure}
To illustrate this point, in Fig.~\ref{fig:lambda}(a), we show the calculated $H_{\rm c2}$ for the same $\lambda (H)$ reproducing the divergence in Ref.~\citen{Ran19}, but with a band Fermi velocity $15\,{\%}$ smaller, or with a $T_{\rm sc}$ $15\,{\%}$ smaller (obtained by reducing the characteristic frequency $\Omega$ by $15\,{\%}$).
In both cases, this is enough to completely suppress the divergence. 
The divergence requires that $\lambda (H)$ increases faster than the suppression of the superconducting state by the orbital limitation.
UTe$_2$ seems to be very close to a perfect balance between both effects. 
As such, the system is certainly very sensitive to many parameters, for example, slight defects or impurities, or a slight change of stoichiometry (there is a solid solution of Te in U around the composition UTe$_2$).
By shifting to the proximity to the FM instability or the change of the band structure, 
the behavior of $H_{\rm c2}$ might be drastically affected.
Let us remark that the field window for the reentrant SC in URhGe shrinks as RRR decreases;
the similar behavior is also observed for the field-reinforced SC in UCoGe.~\cite{Aok19}

A new perspective of UTe$_2$ is the Kondo semiconducting character which emerges in a LDA band structure calculations by
using the structural parameters in Ref.~\citen{Ike06_UTe2}.
As shown in Fig.~\ref{fig:band}, there are very flat bands near the Fermi energy, $E_{\rm F}$, which mainly consist of 5$f$ electrons with $j=5/2$.
At $E_{\rm F}$, a small energy gap ($\sim 130\,{\rm K}$) is realized, indicating the semiconducting nature.
The calculated results do not correspond to the real metallic electronic states at low temperature in UTe$_2$.
Theoretically, to obtain correct sizes of FSs in $f$ electron systems, it is sometimes required to shift the $f$ level in self-consistent LDA calculations.~\cite{Dun09} 
These results suggest that UTe$_2$ is most likely a semi-metal with heavy electronic states,
which may have a very sensitive field-response.

In summary, we confirm SC by resistivity and specific heat experiments as well as a large residual $\gamma$-value below $T_{\rm sc}$. 
The analysis of the resistivity suggests the increase of $\lambda$ with field above $5\,{\rm T}$ for $H\parallel b$-axis, in agreement with the behavior of $H_{\rm c2}$.
It should be clarified if the field response of $\lambda$ in transverse field is dependent on the sample quality.
Compared to the FM superconductors, UCoGe and URhGe, a main issue is to determine if UTe$_2$ remains paramagnetic below $T_{\rm sc}$, or if SC is associated with its non-unitary order parameter to FM ordering.
The divergence of $H_{\rm c2}$ for $H \parallel b$-axis was not observed in our experiments.
However, this may point out its strong sensitivity to sample quality.
The large residual $\gamma$-value below $T_{\rm sc}$ consistently found close to $50\,{\%}$ of the normal phase value. 
This does support the very surprising proposal of a non-unitary state, where only half of the Fermi sea of a given spin direction would be paired. 
This can happen only if spin-orbit coupling is weak enough to avoid inducing SC on the opposite spin Fermi sheets. 
It also requires to induce a FM state on cooling below $T_{\rm sc}$, which could be favored by the closeness to FM instabilities.

\section*{Acknowledgements}
We thank Y.~Tokunaga, S.~Kambe, H.~Sakai, S.~Ikeda, Y. \={O}nuki, K. Ishida, K. Izawa, V. Mineev, and S. Ran
for fruitful discussion.
This work was supported by ERC starting grant (NewHeavyFermion), KAKENHI (JP15H05884, JP15H05882, JP15K21732, JP16H04006, JP15H05745), and ICC-IMR.

\bibliographystyle{jpsj}

\section*{Supplement}

\begin{figure}[tbh]
\begin{center}
\includegraphics[width=0.8 \hsize,pagebox=cropbox,clip]{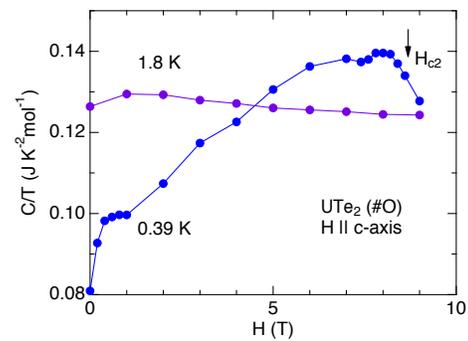}
\end{center}
\caption{(Color online) Field dependence of $C/T$ at $0.39\,{\rm K}$ and $1.8\,{\rm K}$ for $H\parallel c$-axis in UTe$_2$ for sample {\#}O.}
\label{fig:Cp_suppl}
\end{figure}

The field dependence of $C/T$ in UTe$_2$ for $H\parallel c$-axis in the SC state at $0.39\,{\rm K}$ and in the normal state at $1.8\,{\rm K}$ is shown in Fig.~\ref{fig:Cp_suppl}. 
In the SC state at $0.39\,{\rm K}$, which corresponds to $0.25 T_{\rm sc}$,
$C/T$ increases rapidly with field and reaches $0.1\,{\rm J K^{-2}mol^{-1}}$ at $0.4\,{\rm T}$.
It remains almost constant up to $1\,{\rm T}$ and then increases again with a moderate convex curvature.
Further increasing field, $C/T$ starts to decrease because of the collapse of the superconducting state at $H_{\rm c2}\sim 8.6\,{\rm T}$.
Although the temperature is not low enough, the field dependence is not consistent with the isotropic superconducting gap, in which the linear increase of $C/T$ is expected, implying the existence of the anisotropic gap with nodes.
The steep increase at low field may suggest multiband SC,
as is demonstrated in URu$_2$Si$_2$~\cite{Kas07,Kit16}.
In the normal state at $1.8\,{\rm K}$, $C/T$ is almost constant up to $9\,{\rm T}$, in good agreement with the weak field dependence of the resistivity $A$ coefficient, as shown in the main text.

\end{document}